\documentclass[useAMS,usenatbib]{mn2e}
\usepackage{multirow,color}
\usepackage{graphicx}

\newcommand{\Msun}{\ifmmode {M_{\odot}}\else${M_{\odot}}$\fi}
\newcommand{\Rmainz}{\ifmmode {R_{\rm vir,host}(z)}\else$R_{\rm vir,host}(z)$\fi}
\newcommand{\Rmain}{\ifmmode {R_{\rm vir,host}}\else$R_{\rm vir,host}$\fi} 	
\newcommand{\Rvir}{\ifmmode {R_{\rm vir}}\else{$R_{\rm vir}$}\fi}
\newcommand{\Vmax}{\ifmmode {V_{\rm max}}\else{$V_{\rm max}$}\fi}
\newcommand{\Rmax}{\ifmmode {R_{\rm max}}\else{$R_{\rm max}$}\fi}
\newcommand{\lesssim }{{\lower0.8ex\hbox{$\buildrel <\over\sim$}}}
\newcommand{\gessim }{{\lower0.8ex\hbox{$\buildrel >\over\sim$}}}

\title[Identifying Local Group Field Galaxies which have interacted with the Milky Way]{Identifying Local Group Field Galaxies which have interacted with the Milky Way}

\author[M. Teyssier, K. V. Johnston and M. Kuhlen]
   {Maureen ~Teyssier,$^{1}$\thanks{E-mail: maureen@astro.columbia.edu}
   Kathryn V. ~Johnston,$^{1}$ 
   and Michael ~Kuhlen$^{2}$\\
   $^{1}$Department of Astronomy, Columbia University, MC 5246, 550 West 120th Street, New York, NY 10027, USA\\
   $^{2}$Theoretical Astrophysics Center, University of California, Berkeley, CA 94720}

\date{Released 2012 Xxxxx XX}

\begin{document}

\maketitle

\begin{abstract}
We distinguish between Local Group field galaxies which may have passed through the virial volume of the Milky Way, and those which have not, via a statistical comparison against populations of dark matter haloes in the Via Lactea II (VLII) simulation with known orbital histories. Analysis of VLII provides expectations for this escaped population: they contribute 13 per cent of the galactic population between 300 and 1500 kpc from the Milky Way, and hence we anticipate that about 7 of the 54 known Local Group galaxies in that distance range are likely to be Milky Way escapees. These objects can be of any mass below that of the Milky Way, and they are expected to have positive radial velocities with respect to the Milky Way.  Comparison of the radius-velocity distributions of VLII populations and measurements of Local Group galaxies presents a strong likelihood that Tucana, Cetus, NGC3109, SextansA, SextansB, Antlia, NGC6822, Phoenix, LeoT, and NGC185 have passed through the Milky Way.  Most of these dwarfs have a lower HI mass fraction than the majority of dwarfs lying at similar distances to either the Milky Way or M31.  Indeed, several of these galaxies -- especially those with lower masses -- contain signatures in their morphology, star formation history and/or gas content indicative of evolution seen in simulations of satellite/parent galactic interactions. Our results offer strong support for scenarios in which dwarfs of different types form a sequence in morphology and gas content, with evolution along the sequence being driven by interaction history.
\end{abstract}


\section{Introduction}\label{intro}

Dwarfs within the approximate 300 kpc virial radii of the Milky Way and M31 are preferentially small, gas-poor spheroids, compared to their field counterparts which are typically larger, gaseous, and irregularly shaped \citep[e.g.][]{Grebel03,Grcevich09,Weisz11,vandenBergh94}.  This position-morphology relationship, first noted by \citet{Einasto74}, appears universal, as it is found in other galaxy groupings as well \citep[e.g.][]{Skillman03b,Bouchard09}. The position-morphology relationship is attributed to a transformation of gas-rich dwarf irregular galaxies into gas-poor dwarf spheroidals via environmental effects.  That the cumulative environmental effects encountered during a passage through a larger potential are sufficient to transform the morphology of a dwarf is very well motivated by simulations \citep[e.g.][]{Mayer01a,Mayer01b,Kravtsov04,Mayer06}.  

Environmental effects each leave a multitude of signatures on a galaxy.  Tidal stirring has been shown to convert stellar components from disks to bars and finally to pressure supported spheroidal systems \citep[e.g.][]{Klimentowski09}. Shocking and ram-pressure stripping of gas \citep{Sofue94,Grebel03,Mayer10} leaves signatures in the satellite's star formation history, either as starbursts \citep{Hernquist89, Barnes96,Mihos96} or as starvation and quenching of the star formation (see \citet{Kawata08} for a low mass group).  Tidal shock heating is known to disrupt or destroy star clusters \citep{Kruijssen11}. 

Although initially it appeared that these effects might only be highly effective within 50 kpc of a Milky Way-size object \citep{Sofue94,Grebel03}, recent studies (including other effects e.g. tidal effects with UV background \citet{Mayer06}, resonant stripping \citet{D'Onghia09}) show that such a close passage may not be necessary for a morphological transformation.

There are objects that do not fit the rough distance-morphology relationship, because they exist outside the virial radius of the nearest large galaxy, but nevertheless exhibit a morphology that suggests strong interactions (e.g. Tucana).  However, interaction with a Milky-Way-size object is not the only way to affect changes in dwarfs: dwarf-dwarf interactions (or even mergers) have been shown  to stimulate bursts of star formation, and to create irregular morphologies \citep{Mendez99,Bekki08, Besla12}; interactions between dark satellites and dwarf galaxies can also trigger starbursts or a transformation to a spheroidal morphology \citep{Helmi12}; episodic star formation \citep{Gerola80} of the bursty \citep[e.g.][]{Davies88} or quiescent variety \citep[e.g.][]{Tosi92}  has been shown to reduce high gas content and lower metallicity through the interaction of stellar feedback and the interstellar medium; and small galaxies can ionize and blow out (via stellar feedback, and including supernova feedback) enough gas to shut off a star formation episode \citep[e.g.][]{deYoung94, Brinks98}.

Knowledge of the past orbit of a dwarf would be helpful in determining whether prior interaction with the Milky Way is sufficient to explain the properties of objects like Tucana or whether alternative explanations (such as dwarf-dwarf encounters or internal effects) need to be invoked.
Unfortunately, drawing direct, clear connections between the current morphology of an observed object and its past orbit is limited by our observational perspective. It is difficult or impossible to measure more than the angular position, distance and line-of-sight velocity for field dwarfs, and these quantities have been shown to be insufficient to determine a complete, accurate, orbital history for objects in the Local Group \citep{Lux10}.

However, there is precedence for using distance and velocity measurements to draw a connection between morphology and rough orbital history on the larger scale of galaxy clusters. These clusters exhibit a high incidence of so-called ``backsplash galaxies'', defined to be objects on extreme orbits that have taken them through the inner 0.5 \Rvir of a larger potential and subsequently carried them back outside \Rvir. 

 \citet{2005MNRAS.356.1327G} demonstrated in simulations how a population of backsplash galaxies might be probabilistically separated from those infalling to the cluster for the first time using their observed velocities.
Subsequent observations demonstrate that galaxies selected using this approach indeed exhibit unusual or unique morphologies \citep{2010arXiv1012.3114M,2006ApJ...647..946S,2002ApJ...580..164S,2002AJ....124.2440S,2000ApJ...540..113B}.

Owing to the approximately self-similar clustering of dark matter, the research done on clusters provokes questions about the existence and nature of backsplash galaxies on a smaller scale, specifically in the Local Group.  Theoretical work on these scales suggests the existence of satellites on extreme orbits around potentials about the size of the Milky Way.  Around galaxy potentials, \citet{Sales07b} identifies an ``associated'' population of haloes which have at some point passed through the virial volume of the main halo.  Of these, $\sim$6 per cent have apocentric radii greater than 50 per cent of their turnaround radius, and a few have been ejected as far as 2.5 \Rvir.  \citep[Similar populations have also been seen in simulations analysed by][.]{Warnick08,Wang09,Ludlow09,Knebe11a}

Data samples which further inform the extent to which morphology and gas content can be related to dynamical history are growing rapidly.  The study of Local Group objects has recently been invigorated by an influx of new members: SDSS enabled an expansion in the volume probed by star count surveys, which resulted in the discovery of numerous new dwarf satellite galaxies of both the Milky Way and M31 \citep[e.g.][]{Willman05,Belokurov06,Irwin07,Zucker04}.  Moreover, new observational surveys, such as DES \citep{Bernstein11}, SkyMapper \citep{Keller07}, Pan-STARRS \citep{Kaiser02}, and LSST \citep{LSST_ScienceBook_2009,Ivezic08}, will  be even more sensitive to faint magnitude and low surface brightness objects, and are expected to reveal even lower surface brightness objects over even larger volumes of space \citep{Tollerud08}.

Motivated by this confluence of theoretical analyses, recent observational discoveries and promising new surveys, this paper makes connections between dynamically distinct histories for subhaloes seen in a cosmological simulation of structure formation (Via Lactea II, hereafter VLII), and properties of Local Group dwarf galaxies.  More specifically, we establish that it is possible to distinguish field populations which may have passed within the Milky Way-like halo of VLII from those which have not, using observable properties at z=0 (radial distance, line-of-sight velocity and mass).  The $z=0$ distributions of these observable properties for haloes in VLII are given in Section \ref{sec.theory}.  The simulated populations can be used to categorise the orbital histories of Local Group field objects (Section \ref{sec.obs}).  Assuming that morphology is a result of environmental changes over time, we can connect morphology to orbit.  Finally, we discuss whether this rough orbital characterisation provides insight into the morphologies and gas content of nearby field objects in the Local Group (Section \ref{sec.summ}).  The methods we employ, and details of the VLII simulation itself, are described in Section \ref{sec.method}. 

\section{Methods}\label{sec.method}
\setcounter{figure}{0}
\begin{figure*}
\centering
\includegraphics[width=53mm]{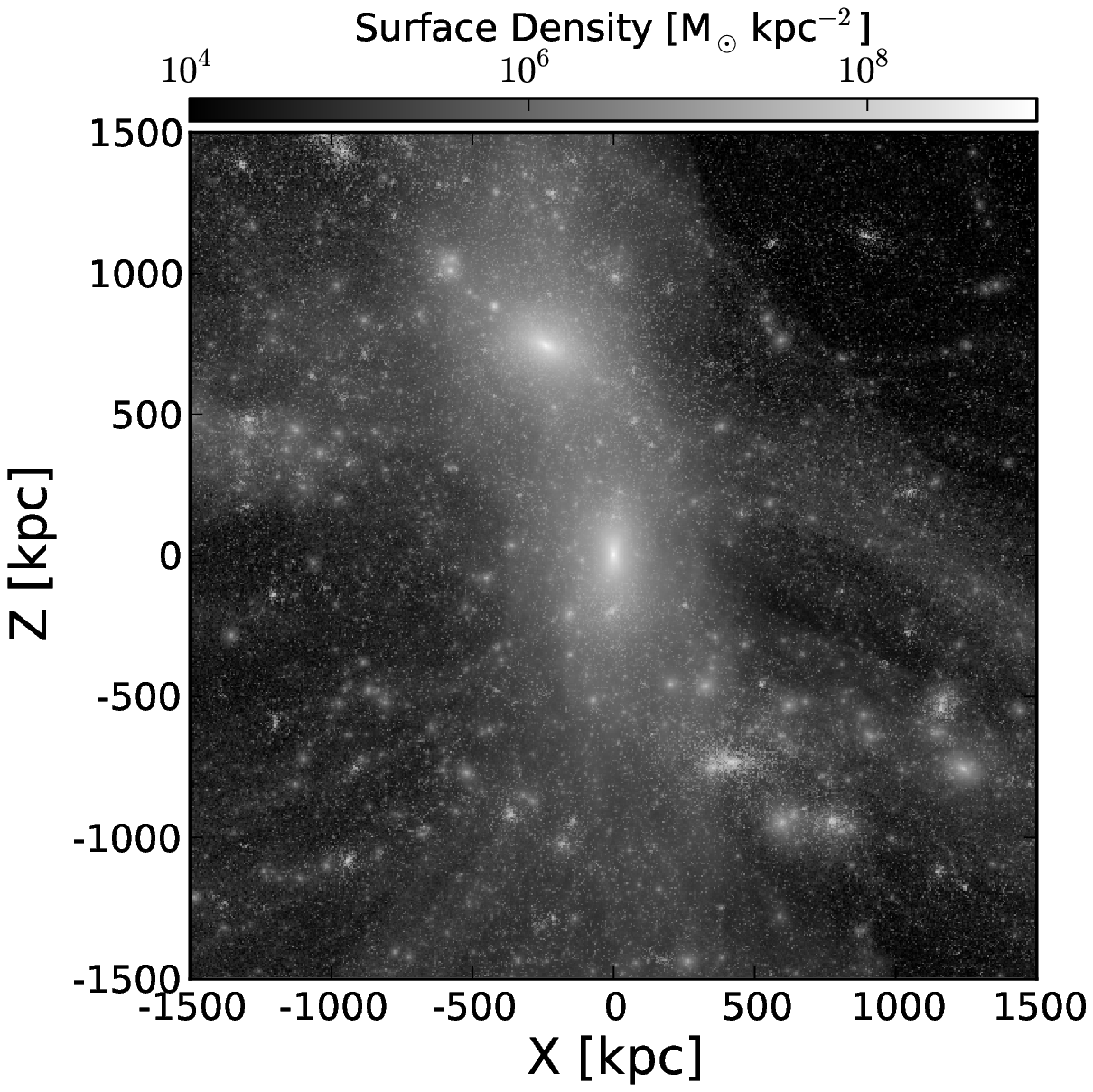}
\includegraphics[width=53mm]{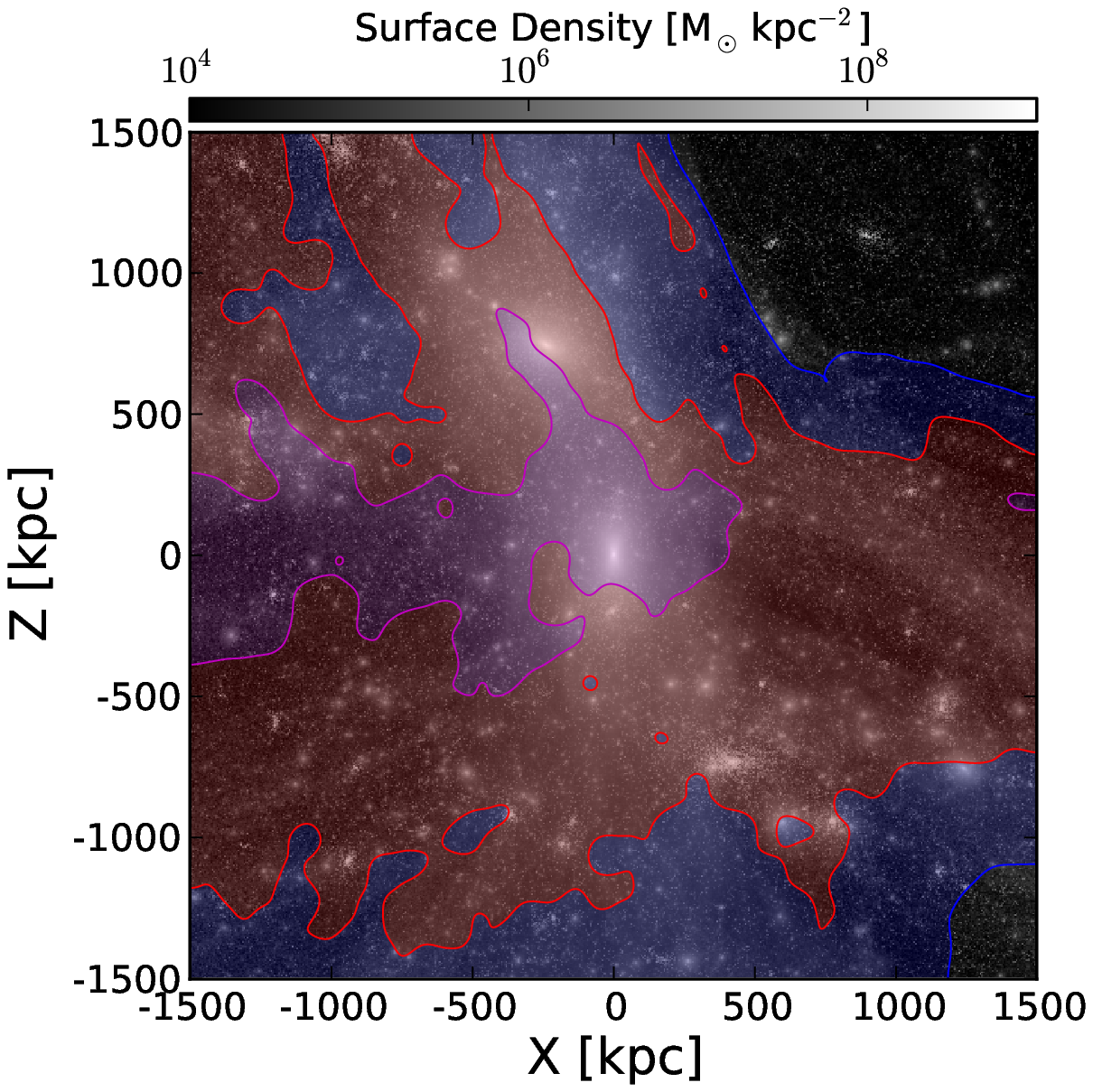}
\includegraphics[width=53mm]{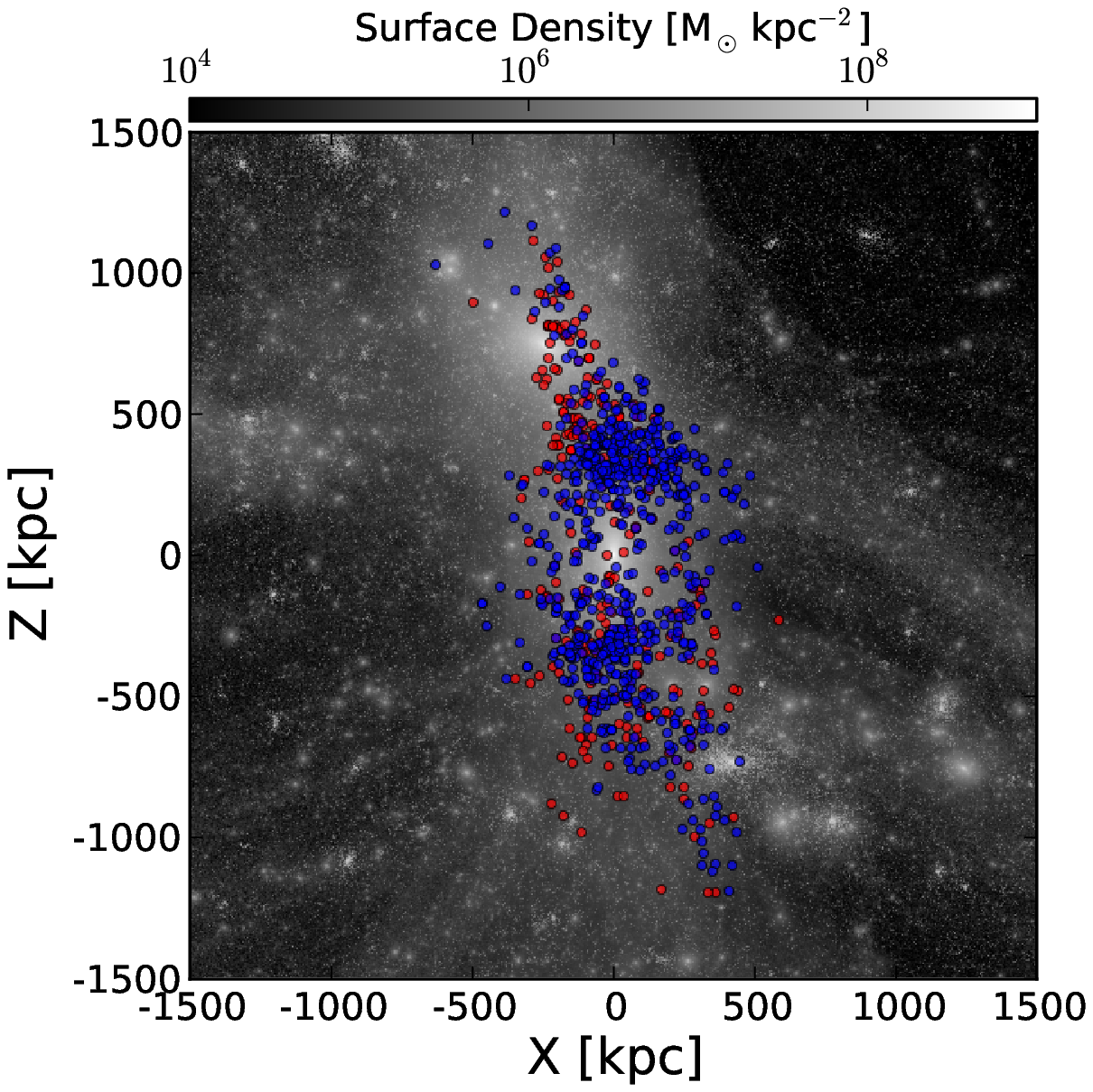}
\caption{The Via Lactea II simulation: A projection of the mass in a 3 Mpc cube onto the XZ plane. Note the central Milky Way size halo, and the less massive halo above and to the left (described in Section \ref{sec.h2}). In the central panel we have over-plotted contours delineating regions containing less than 0.5 per cent (magenta), 5 per cent (red), and 50 per cent (blue) lower resolution particles in projection. In the right panel we have over-plotted the positions of weakly associated (red) and backsplash (blue) haloes.}
\label{fig.vlii}
\end{figure*}

\begin{figure}
\begin{center}
\includegraphics[width=84mm]{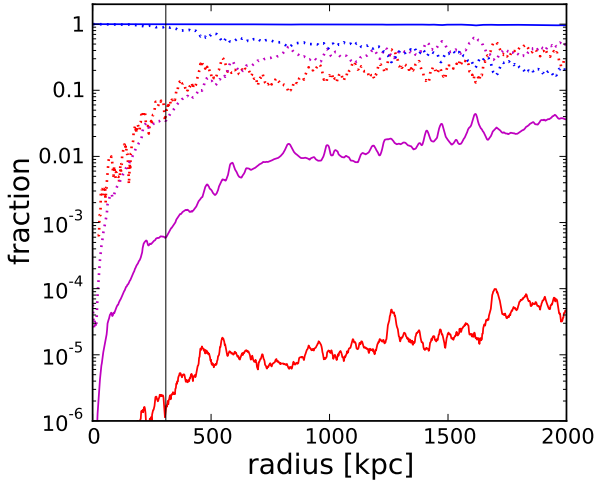}
\caption{The fraction of high-- (blue), intermediate-- (magenta), and low-resolution-- (red) particles contributed to the number of particles (solid lines) and their total mass (dotted) in spherical shells, as a function of distance from the main halo's centre. The main halo's virial radius of 309 kpc is indicated with a solid vertical line.}
\label{fig.pollution}
\end{center}
\end{figure}

VLII is one of the highest resolution cosmological simulation of the formation and evolution of the dark matter halo of a Milky Way like galaxy. The simulation resolves in the initial conditions at $z_i=104$ the Lagrangian region of a halo with a $z=0$ virial\footnote{Here we define virial quantities relative to a density of 92.5 times the critical density  \citep{1998ApJ...495...80B}.} mass and radius of $1.70 \times 10^{12} \Msun$ and $309$ kpc with just over one billion high resolution particles of mass 4,100 \Msun. The surrounding density field is sampled at lower resolution with 29 million and 17 million particles of mass $2.6 \times 10^5 \Msun$ and $1.3 \times 10^8 \Msun$, respectively. The total computational domain of the simulation is (40 Mpc)$^3$. During the evolution, 400 output files, evenly spaced in cosmic time were saved.  \citet{diemand_early_2006} ran the 6DFOF (sub)halo finder on a subset of 27 output files. The $\sim$ 20,000 most massive haloes at $z=4.56$ were linked to their descendant haloes, and their orbits around the host halo traced forward in time. For more detailed information about the VLII simulation and its subhalo population we refer the reader to \citet{diemand_clumps_2008}, \citet{kuhlen_dark_2008}, \citet{madau_fossil_2008}, \citet{zemp_graininess_2009}, and \citet{kuhlen_direct_2012}.

The properties of the VLII simulation make it ideal for our purposes: the small particle mass allows us to follow a large range of halo masses, and trace haloes through order of magnitude changes in mass; the high frequency of outputs allow an accurate assessment of the subhalo interactions with the host halo potential; and the large volume allows us to track subhaloes to large distances beyond the host's virial radius. This last point is one of the distinguishing features of our present work. While previous analyses of the VLII simulation focused on the properties of the subhaloes within the host halo's virial volume, we here consider a population of haloes that at some point passed through the main halo but are found considerably beyond its virial radius at $z=0$. The left panel of Fig.~\ref{fig.vlii} shows a projection along the y-axis of a (3 Mpc)$^3$ region centred on the main host halo at $z=0$. 

However, we exercise caution when using this dataset. Below the galactic scale, baryon and dark matter distributions deviate, due to many of the processes discussed in Section \ref{intro}.  VLII is a purely dark matter simulation, so we use it only to determine the observeables we expect to be independent of baryonic processes on the subgalactic scale --- namely, the location and velocity of galactic-scale objects.  
For example, in simulations which superpose a more realistic matter distribution to represent baryons towards the center of a Milky-Way like object, the number of haloes has been shown to be depleted by about a factor of two within the inner 30kpc of the main halo due to disk shocking \citep{DOnghia10b}.  This destruction takes a few Gyrs.  We analyse a subset of haloes which are found at distances of more than 400 kpc at z=0, a very small number of which would remain within 30 kpc for the required destruction time, so we do not expect this effect to change our results.

\subsection{Subhalo Analysis}

In all Figures, we define a subhalo's mass as
\begin{equation}
M_{\rm Vmax} = {V_{\rm max}^2 R_{\rm max} \over G},
\end{equation}
where \Vmax\ is the maximum circular velocity and \Rmax\ is the radius at which \Vmax\ occurs. This mass is not to be confused with the subhalo's tidal mass or its total gravitationally bound mass. Instead it reflects the mass contained within \Rmax, which is a quantity that for subhaloes is more robustly determined in numerical simulations, but is typically lower than either of the other less well defined masses.

For most of the dark matter haloes in the $z=4.56$ snapshot we were able to identify  any surviving core at $z=0$ by following the orbits derived by \citet{diemand_early_2006}. For a small number of haloes that passed very close to the centre of the main halo we found it necessary to identify the position and velocity of the surviving halo by finding the average location of the particles that were members of  the progenitor object weighted by their $z=4.56$ internal potential energy (i.e. so the derived quantities are biased towards the remaining core). We were then able to match this location to a halo identified by the group finder in the z=0 snapshot. 

\subsection{A Second Host Halo and M31 Analog}\label{sec.h2}
In addition to the main host halo that is the focus of the VLII simulation, a second massive halo (hereafter Halo2) of comparable size to the main halo is apparent in the top left of the projection in Fig.~\ref{fig.vlii}. To obtain the mass of this halo at z=0, we determined the number of bound particles using the potential solver described in \citet{Hernquist92}.  The method begins with the assumption of a basic potential, that is then harmonically modified with the contribution of every particle.  Once the final potential is calculated, unbound particles are discarded.  We iterated this process until the total mass remained constant, to find a total gravitationally bound mass of $6.5 \times 10^{11} \Msun$ with a virial radius of 225 kpc. The distance of the second halo from the main halo is 833 kpc, and they are approaching each other with a speed of 60 km s$^{-1}$.  Overall, we consider Halo2 to be a fortuitous analog to M31, which lies a distance of 785 kpc from the Milky Way \citep{McConnachie05b}, is approaching at 122km s$^{-1}$ \citep{1991trcb.book.....D}, and has a mass of $1.2^{+0.9}_{-0.7} \times 10^{12} \Msun$ \citep{Tollerud11c}.

\subsection{Contamination with Lower Resolution Particles}

At large distances from the main host halo, contamination from lower resolution particles becomes unavoidable. The middle panel of Fig.~\ref{fig.vlii} shows contours delineating regions containing less than 0.5 per cent, 5 per cent, and 50 per cent lower resolution particles by number \textit{in projection}, and Fig.~\ref{fig.pollution} shows profiles of the fraction contributed by high, intermediate, and low resolution particles to the total mass and total number of particles as a function of three-dimensional radius. Throughout our region of interest ($\lesssim 1500$ kpc from the main halo centre) the contamination remains below a few percent by number, but can reach up to almost 50 percent by mass. However, owing to their larger gravitational softening lengths (4.2 and 200 times the high resolution softening length of 40 pc), the dynamical influence of lower resolution particles on highly resolved structures is minimal, and masses, positions, and velocities of such haloes can be accurately determined even in regions subject to non-negligible contamination.

\subsection{Subhalo Nomenclature}\label{sec.nom}

To examine the relationship between orbital histories and $z=0$ mass, radial distance and velocity, we separate the haloes in the VLII simulation into basic categories based on whether they have passed deeply, shallowly or not at all through the virial radius of the main halo. We employ the following commonly used nomenclature for these categories:
\begin{description}
\item [ASSOCIATED:] Haloes which have passed within half the virial radius of the main halo, and exited by z=0, are `\textbf{backsplash}' haloes.  Haloes which have only passed through outskirts of the halo (within 0.5-1 virial radius), and exited by z=0,  are `\textbf{weakly associated}'.
\item [UNASSOCIATED:] Haloes which remain outside the virial radius of the main halo to z=0 are `unassociated' haloes.  
\item [SUBHALOES:] Haloes found within the virial radius at z=0, we simply call ``subhaloes''.  
\end{description}

\section{Results I: Distribution of subhaloes in VLII} \label{sec.theory}

In this Section we examine the VLII halo population to determine if there are observable differences between their orbital history categories.

\subsection{Halo Category Statistics}\label{subsec.pop}

\begin{table}
\begin{center}
\begin{tabular}{lcc}
  \multicolumn{3}{c}{Statistics of VLII Halo Categories} \\  \hline \hline
  \multirow{2}{*}{Category} & Remaining & Ever entered \\
                            & at $z=0$  & the main halo\\
   \hline  \hline
   Backsplash  & 695 & 5,352 \\
   \hline 
   Weakly Associated  & 312 & 647  \\
   \hline 
   Subhaloes & 1,534 & 5,999 \\
   \hline 
   Destroyed & 3,458 & 5,999 \\
   \hline
   Unassociated & 7,513  & -- \\
   \hline 
\end{tabular}
\caption{Halo counts for the most massive (presumably star-forming) halo categories in VLII.  Categories are defined in Section \ref{sec.nom}.}
   \label{t2}
\end{center}
\end{table}

Since we are interested in observable results, we eliminate from the VLII halo catalogue haloes that were not massive enough to allow for gas to condense and star formation to occur. For this purpose we reject haloes that never reach a mass of $M_{\rm Vir}(z) > 10^7 \Msun$, similar to the approach taken in \citet{Rashkov11}. There are 13,512 haloes above this mass cut, and these are the only haloes we consider in the following analysis.

Of the 13,512 massive haloes, 5,999 (44 per cent) are at some point found within the redshift-dependent virial radius, \Rmainz, and the majority of these (5,352, 89 per cent) deeply penetrate the main halo, passing within half \Rmainz.  A small fraction of the deeply penetrating haloes (695, 13 per cent) are found outside \Rmain\ at z=0, and are therefore `backsplash' haloes. Additionally, 647 haloes pass through the host's virial volume, but never enter the central 0.5 \Rmainz.  A larger fraction of the shallowly penetrating haloes, almost half (312), make their way back outside \Rmain\ by z=0 to become `weakly associated' haloes.  

The majority of haloes that pass within \Rmainz are completely destroyed and have no identifiable $z=0$ remnant (3,458, 58 per cent).  Only about a quarter (1,534) of haloes survive within \Rmain to z=0, and are thus `subhaloes'.  (The remaining 1007, or 17 per cent are the weakly associated and backsplash haloes.)  There are also 7,513 haloes in our catalogue that never enter the main halo's virial volume at all and are hence `unassociated' haloes not likely to have been affected by the main halo.  These halo statistics are summarised in Table~\ref{t2}.  

The fraction of associated haloes to total simulation haloes we find (10 per cent) is slightly larger than the 9-4 per cent quoted for increasing halo masses in \citet{Wang09}, despite simulation differences.  There are several plausible explanations for this difference.  VLII's analysis focuses solely on the high resolution area around two haloes between $10^{11}$ and $10^{12}$ \Msun, while analysis in \citet{Wang09} covers $\ge$ 22,000 haloes in that mass range in a (100$h^{-1}$ Mpc)$^3$ volume.  Hence, their value of 9-4 per cent is a very robust average, whereas our system could be an outlier due to, perhaps, the proximity of our two main haloes.  \citet{Wang09} also find that the fraction of associated haloes decreases with increasing satellite halo mass. This trend, in combination with our ability to trace much lower satellite halo masses (the particle mass in \citet{Wang09} is $6.2 \times 10^8$ \Msun), likely accounts for our slightly higher associated fraction.

We now volume-limit the z=0 halo results to make them more readily comparable to the Local Group sample examined in the remainder of the paper.  Taking the haloes within 1.5 Mpc ($\sim$5 \Rmain) decreases the number of unassociated haloes from 7,513 to 6,888. Of the potentially star-forming subhaloes (corresponding to theoretically predicted dwarf galaxies) found between 1-5 virial radii (see Subsection \ref{subsec.dist} for justification of radius limit) at $z=0$, $\sim$13 per cent have passed within the virial radius of the main halo during their history.  Considering that there exist at least 54 Local Group galaxies in this radius range, we expect that $\sim$7 of these are examples of associated galaxies that have passed within the virial volume of the MW in the past.

\subsection{Host Halo Membership Subhaloes at $z=0$}\label{member}
We briefly discuss the membership of subhaloes at z=0, as defined by an orbital energy calculation with respect to either the Milky Way-like main halo or Halo2.

Note that, despite being outside the virial radius (and in some cases {\it far} outside) most of the associated subhaloes are still gravitationally bound to the main halo at z=0.  A minority of the backsplash and weakly associated subhaloes, 7 per cent and 17 per cent respectively, have become unbound from both the main halo and Halo2.  Remarkably, a small fraction of the associated haloes, 5 per cent and 4 per cent of strongly and weakly associated respectively, are bound to Halo2 but not to the main halo, and thus appear to have been captured by Halo2, making these objects so-called ``renegade haloes'' \citep{Knebe11a}.

We note that Halo2, the second largest halo in the VLII simulation, has it's own associated, unassociated and subhalo populations.  Apart from the renegade haloes mentioned above, these are encompassed within the main halo's unassociated population. Our analysis could be duplicated from the perspective of Halo2, and those results would be particularly interesting if transverse velocities of more Local Group Field objects were known.

\subsection{Spatial Distributions}\label{subsec.dist}

\begin{figure}
\begin{center}
\includegraphics[width=84mm]{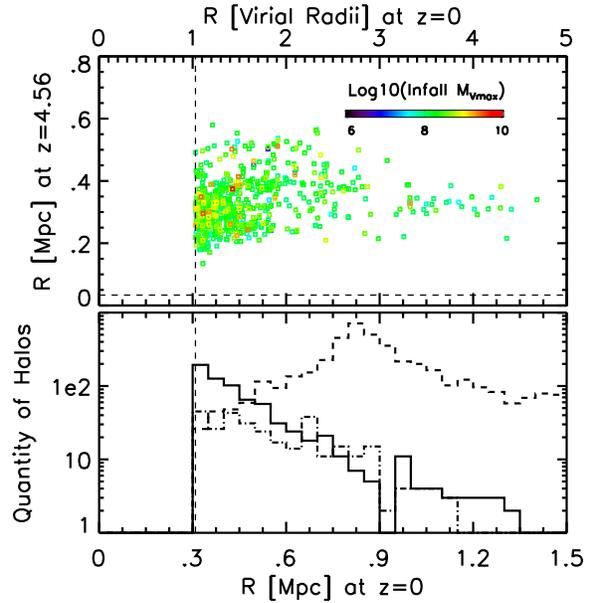}
\caption{TOP: A comparison of the radial distance from backsplash haloes to the central, most massive halo at redshift z=4.56, and redshift z=0.  Backsplash haloes are coloured with the log of their infall mass.  Distances are shown in Mpc and in virial radii of the main halo.  The virial radius of the halo is shown at redshift z=0 (vertical, dashed line).   Haloes are scattered to 5 virial radii.  BOTTOM:  Quantity of haloes as a function of z=0 radial distance for backsplash (solid), weakly associated (dot-dashed) and unassociated haloes (dashed).}
\label{fig.distance}
\end{center}
\end{figure}

Figure \ref{fig.distance} shows the radial distance from VLII haloes to the centre of main halo compared between z=4.56 and z=0. Weakly associated haloes have been scattered out to 1.5 Mpc ($5 \times \Rvir(z=0)$), and backsplash haloes are found past 1.2 Mpc at z=0. The backsplash haloes are plotted as squares and are colour-coded by infall mass, the mass that they had just prior to their first crossing of \Rmainz.
A trend of decreasing infall mass with distance from the host halo, as may be expected from multi-body interactions with the host halo, is not readily apparent (see \S\ref{subsec.mass}). Histograms of the $z=4.56$ and $z=0$ distances for the two populations are shown in the left and bottom panels of the figure.  We find that the radial distribution of associated (weakly + backsplash) haloes is well fit by a simple power law: $dN/dR \propto R^{-3.7}$.

It is somewhat surprising that associated haloes are found as far out as 5 \Rvir. The analytic analysis of a cluster halo by \citet{2004A&A...414..445M} should roughly scale down to a galaxy size halo, so backsplash haloes in both cluster and galaxy simulations should only be found out to $\sim$2.5 \Rvir\ at z=0. This was seen in simulations of isolated galaxy potentials \citep{Sales07b}.  However, \citet{Wang09} and \citet{Ludlow09} have also found associated haloes to large distances, 4$R_{200}$ and 5$R_{200}$, respectively. As shown in the right panel of Fig.~\ref{fig.vlii}, the associated haloes in VLII fill an elongated volume of space, oriented toward Halo2. Perpendicular to the elongated axis, associated haloes are found out to only 2.5 \Rmain, approximately the value predicted by \citet{2004A&A...414..445M} for an isolated halo.  

It appears that the unexpectedly large radial extent of associated haloes in VLII may be due to the strong anisotropy of cosmological infall and the presence of Halo2, which itself is still infalling along one of the three main filaments feeding the main halo.  This could also account for the large extent of the associated objects found in \citet{Wang09,Ludlow09}. (Assuming that some of the parent galaxies in \citet{Ludlow09} had strong filaments or a companion in the 1-2Mpc range, beyond their isolation criteria of 1Mpc.)

\subsection{Mass Distributions}\label{subsec.mass}

\begin{figure}
\begin{center}
\includegraphics[width=84mm]{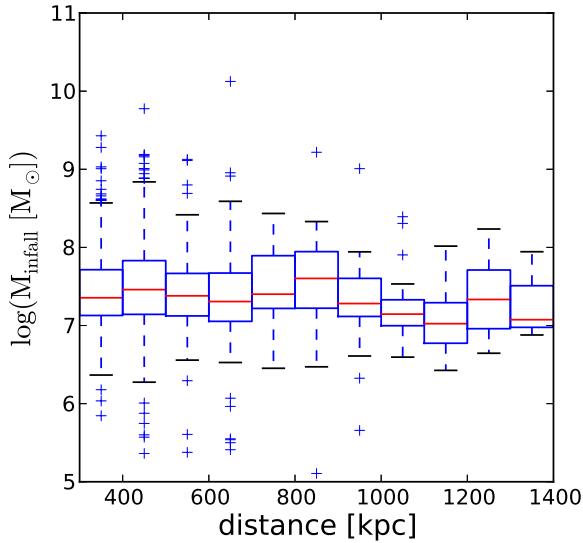}
\caption{Distributions of $\log(M_{\rm infall})$ of the associated haloes as a function of their $z=0$ distance from the main halo, in bins of 100 kpc width. The median of $\log(M_{\rm infall})$ is given by the red line, the box extends from the 25$^{\rm th}$ to the 75$^{\rm th}$ percentile, the whiskers (dashed lines) from the 12.5$^{\rm th}$ to the 87.5$^{\rm th}$ percentile, and haloes outside this range are shown as individual crosses. There is no evidence for an inverse relation between $M_{\rm infall}$ and distance.}
\label{fig.Min_vs_D}
\end{center}
\end{figure}

Previous work on the ejection of subhaloes from a galaxy potential considered a slingshot effect \citep{2007MNRAS.382.1901S} and a tidal impulse \citep{Teyssier09}. \citet{2007MNRAS.382.1901S} examined the origin of the two most dynamically extreme objects, one with a z=0 distance of 2.5$R_{vir}$ and the other with a velocity of 2$V_{vir}$, and found both to have originated from pairs of objects, one of which received an energetic impulse from the host potential at pericentre.  \citet{Teyssier09} describe the same mechanism in a different way, focusing on a distribution of objects (a satellite or a group of satellites) instead of a pair of objects (see also \citet{Ludlow09}).

Both of these mechanisms predict an inverse correlation between mass and distance, because the smaller, more peripheral members of an infalling group experience the largest energy gain during the group's pericentre passage.  We expect to see this signature if these mechanisms are solely responsible for the associated halo population in the VLII simulation.  We split the associated halo sample by $z=0$ distance into bins of 100 kpc width, and look at the distributions of $\log(M_{\rm vmax})$ at infall, within each bin (Fig.~\ref{fig.Min_vs_D}).  Neither the median of the distributions nor its scatter exhibit any noticeable trends with distance.  We don't view this absence of inverse correlation between mass and distance to be strong evidence that the mechanisms described above aren't occurring.  Rather, it indicates that other dynamical processes also have a comparable effect on the ejection of subhaloes to beyond the virial radius.  These other processes could include dynamical interactions that occur on multiple levels, in conjunction with the main potential i.e. subhalo-subhalo, subhalo-group, or group-group interactions.  Regardless, our analysis indicates that one should not expect a mass-distance bias in the associated dwarfs around the Milky Way.

\subsection{Velocity Distributions}\label{subsec.vel}

\begin{figure}
\begin{center}
\includegraphics[width=84mm]{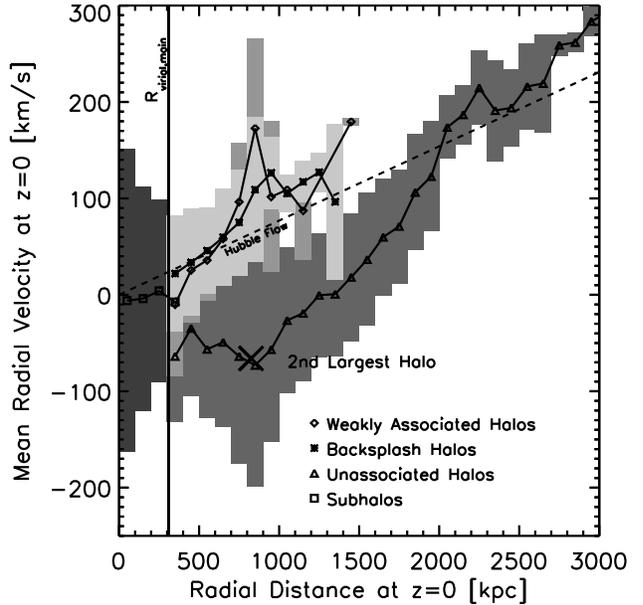}
\caption{The mean radial velocity per radial distance bin for four populations of haloes: subhaloes remaining within the virial radius of the central halo at z=0 (squares), haloes which have never entered the virial radius of the central halo (triangles), haloes which have passed within 0.5-1 virial radii of the central halo (diamonds),  and haloes which have previously passed within 0.5 virial radii of the central halo (star), are shown on the radial distance vs. radial velocity plane at redshift 0. The 68 per cent confidence region for the subhalo, unassociated, weakly associated and backsplash populations is shown in darkest grey, dark grey, medium grey and light grey, respectively.  The solid vertical line is the virial radius of the central halo at z=0, and the dashed line shows the Hubble Flow. }
\label{fig.veldist}
\end{center}
\end{figure}

\begin{table*}
\begin{center}
\begin{tabular}{lccccll}
  \multicolumn{7}{c}{Properties of Local Group Objects} \\  \hline \hline
Common Name(s) & $D_{\rm helio}$ & $V_{\rm helio}$ & $D_{\rm gsr}$ & $V_{\rm gsr}$ & Class & Sources\\ 
  & [kpc] & [km s$^{-1}$] &  [kpc] & [km s$^{-1}$]  & &  \\ \hline

Phoenix Dwarf     & 406.  & 56.   & 401.  & -37.  & dIrr/dSph & \citet{Hidalgo09,Cote97} \\
LeoT              & 415.  & 35.   & 421.  & -69.  & dIrr/dSph  & \citet{deJong08,Irwin07} \\
NGC6822 (DDO209)  & 489.  & -57.  & 486.  & 57.   & Irr  & \citet{Wyder03,Irwin07} \\
IC10              & 715.  & -348. & 711.  & -137. & dIrr  & \citet{Kim09,Huchra99} \\
IC1613 (DDO8)     & 748.  & -234. & 740.  & -150. & Irr & \citet{Rizzi07,Lu93} \\
LGS3              & 769.  & -287. & 762.  & -146. & dIrr/dSph  & A, \citet{Huchtmeier03} \\
Cetus             & 755.  & -87.  & 747.  & -23.  & dSph  & A, \citet{Grcevich09}  \\
LeoA (DDO69)      & 809.  & 24.   & 815.  & -21.  & dIrr  &  \citet{Tammann08,Huchtmeier03} \\
Tucana            & 890.  & 194.  & 887.  & 96.   & dSph/dE4 &  \citet{Bernard09,Fraternali09} \\
Aquarius (DDO210) & 1071. & -141. & 1066. & -12.  & dIrr/dSph  & \citet{Karachentsev02,Koribalski04} \\
WLM (DDO221)      & 966.  & -122. & 958.  & -57.  & dIrr & \citet{Gieren08,Koribalski04} \\
SagDIG            & 1040. & -79.  & 1037. & 20.   & Irr & \citet{Karachentsev02,Koribalski04} \\
Pegasus (DDO216)  & 1070. & -183. & 1062. & -9.   & dIrr/dSph  & \citet{Meschin09,Huchtmeier03} \\
Antlia Dwarf      & 1290. & 362.  & 1296. & 137.  & dIrr/dSph  & \citet{Dalcanton09,Huchtmeier03} \\
NGC3109 (DDO236)  & 1260. & 403.  & 1266. & 179.  & Irr/bar & \citet{Dalcanton09,Lauberts89}  \\
SextansA (DDO75)  & 1380. & 324.  & 1387. & 155.  & dIrr & \citet{Dalcanton09,Koribalski04} \\
SextansB (DDO70)  & 1390. & 300.  & 1397. & 157.  & dIrr & \citet{Dalcanton09,Huchtmeier03} \\
VV124 (UGC4879)   & 1360. & -29.  & 1364. &  16.  & dIrr/dSph  & \citet{Bradley11,Kirby12}\\
\hline
M31               & 785.  & -300. & 779.  & -110. & SA(s)b & A, \citet{1991trcb.book.....D} \\
AndXVI            & 525.  & -367. & 518.  & -201. &  dSph? & \citet{Ibata07,Letarte09}\\
NGC185            & 616.  & -202. & 611.  &   -3. & dSph/dE3p  & A, \citet{Bender91} \\
AndII             & 652.  & -188. & 645.  &  -29. & dSph &  A  \\
NGC147 (DDO3)     & 675.  & -193. & 670.  &    9. & dSph/dE5  & A, \citet{Yang10} \\
AndXIV            & 735.  & -481. & 728.  & -316. & dSph? & \citet{Majewski07}\\
AndI              & 745.  & -368. & 739.  & -185. & dSph & A  \\
AndIII            & 749.  & -314. & 742.  & -128. & dSph & A, \citet{Karachentseva98}    \\
AndX              & 760.  & -164. & 754.  &   19. & dSph &\citet{Zucker04} \\
AndVII            & 763.  & -307. & 758.  &  -81. & dSph & A, \citet{Karachentsev01} \\
AndIX             & 765.  & -209. & 759.  &  -22. & dE & A, \citet{Zucker04} \\
AndXV             & 770.  & -323. & 764.  & -154. & dSph? & \citet{Ibata07,Letarte09}\\
AndV              & 774.  & -397. & 769.  & -212. & dSph  & A, \citet{Mancone08} \\
AndXXII           & 794.  & -127. & 787.  &   14. & dSph? & \citet{Martin09,Tollerud11c}\\
M32 (NGC221)      & 817.  & -200. & 811.  &  -11. & cE2  &  \citet{Fiorentino10} \\
NGC205 (M110)     & 824.  & -241. & 818.  &  -50. & dSph/dE5 & A, \citet{Bender91} \\
AndXII            & 830.  & -525. & 823.  & -349. & dSph? & B, \citet{Chapman07,Tollerud11c}\\
AndXXI            & 859.  & -362. & 853.  & -151. & dSph? & \citet{Martin09,Tollerud11c}\\
AndXI             & 870.  & -462. & 763.  & -286. &  dSph? & B, \citet{Tollerud11c}\\
AndXIII           & 880.  & -185. & 873.  & -13.  &  dSph? & B, \citet{Yang11}\\
M33 (NGC598)      & 884.  & -179. & 877.  &  -36. & SA(s)cd & \citet{Martin09} \\
AndXVIII          & 1355. & -332. & 1349. & -121. &  dSph? & \citet{McConnachie08,Tollerud11c}\\
\hline
\end{tabular}
\caption{Choice of distance and velocity are shown in both heliocentric and galactocentric reference frames.  Details for reference frame conversion are found in \S\ref{sec.obs}.  In the sources column, A refers to \citet{McConnachie05b} and B refers to \citet{Martin06}.}
   \label{t3}
\end{center}
\end{table*}

In Fig.~\ref{fig.veldist} we show the means and standard deviation of the radial velocity distributions in the same 100 kpc $z=0$ distance bins, for each of our four subhalo categories: backsplash, weakly associated, unassociated haloes, and subhaloes (with increasingly darker shades of Gray). Beyond the virial radius, and up to 1.5 Mpc, the unassociated haloes are inflowing (negative radial velocity), either onto the main halo, or onto Halo2, which we have marked with an X on the figure. The weakly associated and backsplash haloes are outflowing with approximately the Hubble Flow (diagonal dashed line) or even higher radial velocities.  \citet{Wang09} also finds a velocity offset for the associated population.  On average VLII associated populations have radial velocity that is 50 - 100 km s$^{-1}$ higher than the unassociated haloes.  This offset is much larger than observational uncertainties in velocity.  These results raise the exciting possibility of using this radial velocity signature to observationally assign a likelihood of being an associated halo, to actual Local Group dwarf galaxies. In \S\ref{sec.obs} we do exactly that, by comparing the locations of real Local Group dwarf galaxies in the $v_r-r$ plane with the predictions from the VLII simulation.

Our confidence in assigning associated halo status to observed dwarf galaxies hinges on how well separated the associated and unassociated populations are in the $v_r-r$ plane. Around $d \sim$850 kpc the presence of Halo2 leads to an increase in the radial velocity dispersions in both halo categories, such that the distributions significantly overlap. However, spatial information can be used to increase the distinction between the populations.  Selecting haloes with large angular separation from the centre of Halo2 (from a vantage point at the centre of the main halo) significantly decreases overlap in the radial velocity distribution between associated and unassociated haloes, and leads to a more appropriate analysis for some Local Group objects that also lie at large angular separation from Halo2.

There is no significant difference between the $v_r-r$ distributions of backsplash and weakly associated haloes, making it impossible to distinguish between weakly associated and backsplash haloes with this method.  Unfortunately, we cannot thereby separate objects we expect to have undergone more dramatic changes in their morphology (backsplash) from those with relatively more minor transformations (weakly associated). Note that this distinction may be less important in the resonant stripping model proposed by \citet{D'Onghia09}, in which heavy stripping and morphological transformation can occur even for subhaloes without close pericentre passages, provided that they enter the host halo on a retrograde orbit.

\section{Comparison of Simulation Results to Observations}\label{sec.obs}

To briefly recap the main results of \S\ref{sec.theory}, from an analysis of the subhalo population in the VLII simulation, we expect that: (i) $\sim$13 per cent of the Local Group field dwarfs have passed through the virial volume of the Milky Way; (ii) these associated dwarfs can be found out to 5 \Rvir\ ($\approx 1.5$ Mpc); (iii) the associated dwarf population does not necessarily exhibit any strong trends in mass with distance; (iv) associated dwarfs are likely to have positive radial velocities with respect to the Milky Way, of order or greater than the Hubble Flow, and in contrast to unassociated haloes which typically have negative radial velocities out to $\sim 1.5$ Mpc; and lastly (v) it's possible that there are so-called renegade satellites around M31, i.e. MW escapees that have become bound to M31. 

In the following, we identify Local Group field objects which may be associated with the Milky Way by comparison of their dynamical properties with those of the populations in Via Lactea II.  We then augment our argument for the plausibility of their association by including the observed properties of the objects, including stellar population ages, and  gas content. 

\subsection{Radial Distance and Velocity Comparison}

\begin{figure*}
\begin{center}
\includegraphics[width=160mm]{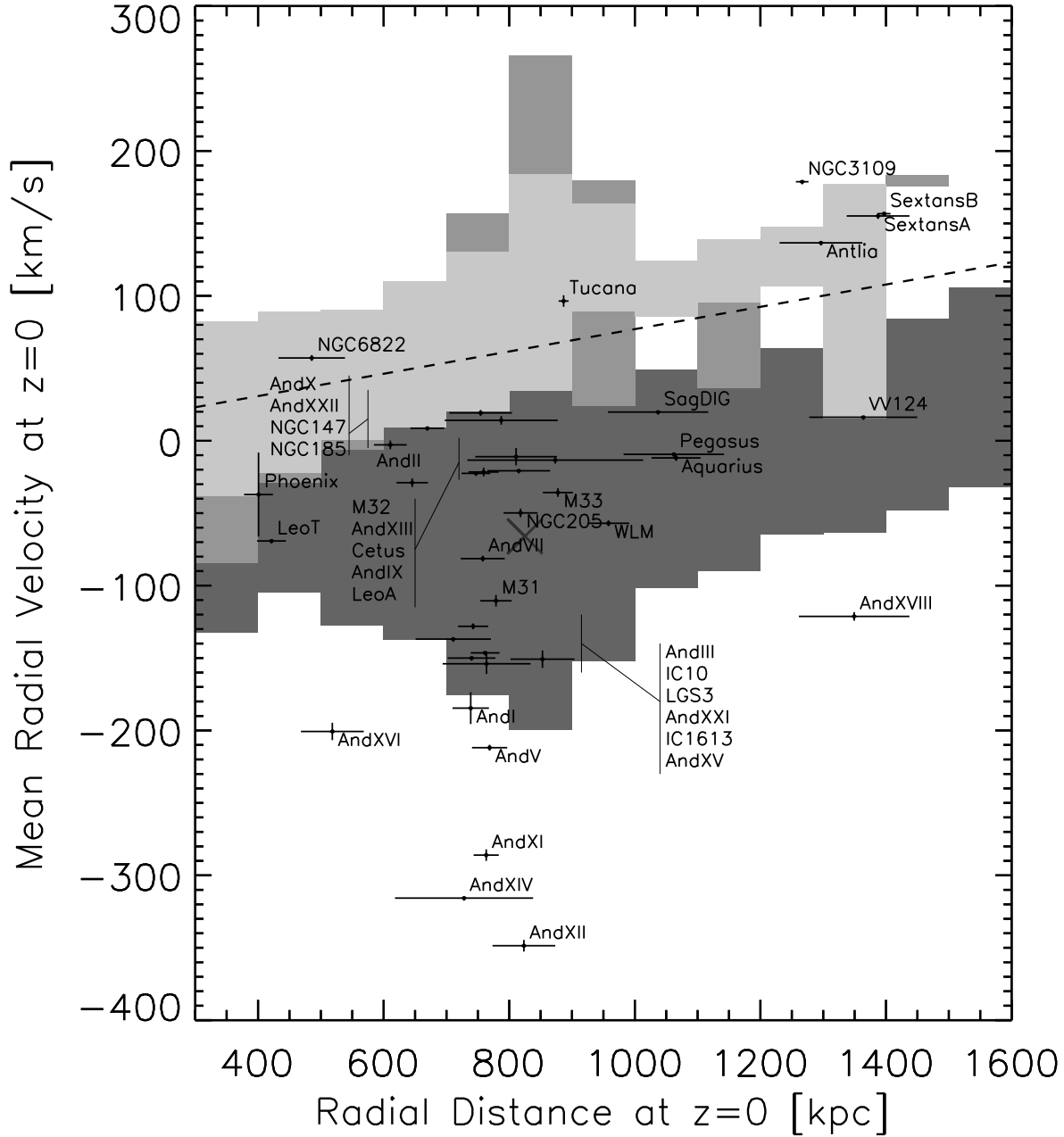}
\caption{The observed radial distances [kpc] and velocities [km s$^{-1}$] in the galactocentric frame are over-plotted on the 1-sigma distributions of the simulated halo populations from VLII (weakly associated, backsplash, unassociated, and subhalo in increasingly dark shades of grey).  Note that of the galaxies with large negative velocity (less than -200 km/s) all (except AndXVIII)  lie within the full distribution of the VLII haloes.  Moreover, M31's actual mass could be up to a factor of 2 larger than that of our M31-analogue. It's true velocity dispersion thus could be a factor of $sqrt(2)$ larger, accounting for the dwarfs that fall on the lower edges of the VLII distribution. }
\label{fig.mw_obs_err}
\end{center}
\end{figure*}

\begin{table}
\begin{center}
\begin{tabular}{llllll}
\multicolumn{6}{c}{Fractional Likelihood of being ``associated'' to the Milky Way} \\  \hline \hline
Name     & \multicolumn{3}{l}{Likelihood per bin (Haloes per bin)}  & Morphology & Mass \\ 
         &total pop. &  45$^\circ$ away & 90$^\circ$ away             &            & [$10^6 M_{\odot}$] \\ \hline \hline
NGC3109  & 1.00 (2)  & 1.00 (2)  & 1.00 (2)  & Irr & 6550 \\
NGC6822  & 0.64 (50) & 0.87 (15) & --        & Irr & 1640\\
SextansB & 1.00 (3)  & 1.00 (3)  & 1.00 (3)  & dIrr & 885 \\
SextansA & 1.00 (3)  & 1.00 (3)  & 1.00 (3)  & dIrr & 395 \\
NGC185   & 0.56 (84) & --        & --        & dSph/dE3p & 130\\
Phoenix  & 0.70 (77) & 0.70 (56) & --        & dIrr/dSph  & 33\\
Antlia   & 1.00 (2)  & 1.00 (2)  &  1.00 (2) & dIrr/dSph  & 12 \\
Leo T    & 0.70 (77) & 0.70 (56) & 0.74 (38) & dIrr/dSph  & 8\\
Tucana   & 0.36 (22) & 1.00 (2)  &  1.00 (2) & dSph & ? \\
Cetus    & 0.17 (95) & 0.60 (27) & --        & dSph & ? \\
\hline
M33      & 0.01 (110) & --        & --        & SA(s)cd & $5 \times 10^4$\\
NGC147   & 0.17 (95)  & --        & --        & dSph/dE5 & 110\\
LeoA     & 0.02 (161) & 0.17 (18) & 0.17 (18) & dIrr & 80\\
Pegasus  & 0.02 (47)  & --        & --        & dIrr/dSph  & 58\\
Aquarius & 0.02 (47)  & 0.00 (27) & --        & dIrr/dSph  & 5\\
AndII    & 0.05 (60)  & --        & --        & dSph & ? \\
AndX     & 0.02 (161) & --        & --        & dSph & ? \\
AndIX    & 0.02 (161) & --        & --        & dSph & ? \\ 
AndXXII  & 0.02 (161) & --        & --        & dSph & ? \\
  \hline
\end{tabular}
\caption{The fractional likelihood that the object is 'associated' (passed through the virial radius of the Milky Way) is defined by comparison of the galactocentric velocity and galactocentric distance of Local Group objects to VLII data.  Analysis is repeated for objects with angular separation of more than $45^{\circ}$ and more than $90^{\circ}$.  These additional analyses use subsets of the VLII populations which fit there same angular constraints.  Objects below the horizontal line have less than 50 per cent likelihood that they are associated with the Milky Way.  Total mass values are taken from \citet{Mateo98}, or from \citet{Brown07} for LeoA, \citet{Corbelli03} for M33, and \citet{Simon07} for LeoT.}
\label{t4}
\end{center}
\end{table}

As discussed in \S\ref{subsec.vel}, the separation between the associated and unassociated populations from VLII in the $v_r-r$ plane makes it possible to use these same properties of Local Group field objects to predict the likelihood that they are either associated or unassociated with the Milky Way.  Velocity and distance measurements of Local Group objects with sources are summarised in Table \ref{t3}. Errors in measurement are as reported by the source, or as found in NED. The distances and velocities in Table \ref{t3} are converted from the heliocentric reference frame to the galactocentric reference frame, for comparison to VLII data.  The following assumptions are made: the Solar System lies at a distance of 8.3 kpc from the galactic centre \citep{Gwinn92}.  The local rotation speed is $\Theta_{0}$=236 km s$^{-1}$; the speed of a closed orbit at the position of the Sun relative to the Galactic center \citep{Bovy09}.  The relative motion of the Sun is ($U_\odot$,$V_\odot$,$W_\odot$)=(11.1,12.24,7.25) km s$^{-1}$ \citep{Schonrich10}.

Fig. \ref{fig.mw_obs_err}  repeats Fig. \ref{fig.veldist}  with the Local Group data over-plotted. It is clear from this Figure that there are several examples of field objects in the Local Group that fall in the region outlined by associated haloes in VLII and hence are likely to have interacted with the Milky Way some time in the past.   The $\sim 50-100$ km s$^{-1}$ separation between the objects that are obviously bound to the second massive (Andromeda-like) halo and those objects found above the Hubble Flow is much larger than known observational uncertainties. 

For a more quantitative determination of whether an object is likely to be associated with the Milky Way, we divide the radial velocities and distances of the associated and unassociated VLII halo populations into bins of size 100 kpc and 50 km s$^{-1}$. The fraction of haloes in a $v_r-r$ bin that are ``associated'' gives a rough estimate  of the likelihood of an observed Local Group dwarf in the same bin having interacted at some time in the past with the Milky Way.  These likelihoods are listed in Table \ref{t4} for our most likely associated halo candidates.

Note that the simple radial distance and velocity test does not take into account the full spatial distribution of Local Group objects.  In Figure \ref{fig.mw_obs_err}, several observed objects lie in regions of the $v_r-r$ plane where the wings of the radial velocity distributions of associated and unassociated populations overlap due to the presence of Halo2 (as discussed in \S\ref{subsec.vel}).  However, some of these objects lie at large angular separations from M31 (e.g. Tucana). To address this issue we also performed comparisons of the most isolated dwarfs (more than $45^\circ$ and more than $90^\circ$ from Andromeda) against the VLII distributions for all haloes more than $45^\circ$, or $90^\circ$ respectively, from Halo2. These corrected likelihood estimates are included in Table \ref{t4}.

From Table \ref{t4} we expect that the following Local Group Objects have with high likelihood ($>$50 per cent) at some point in time passed through the virial radius of the Milky Way: NGC3109, SextansA, SextansB, Antlia, Cetus, Tucana, NGC6822, Phoenix, LeoT, and NGC185.  Note that the zero-velocity radius of the Local Group is 0.96 Mpc \citep{Karachentsev09}.  This radius cut-off has been used in the past to exclude the Antlia Group (Antlia, NGC3109, SextansA and SextansB) from membership in the Local Group of Galaxies \citep{Courteau99}.  If these objects are not currently members, our results indicate that they were likely to be in the past. The rest of the objects found in Table \ref{t3} have a likelihood of association with the Milky Way that is very low or zero.  The fractional likelihood that the following objects are associated is less than 1 per cent: IC10, IC1613, LGS3, WLM, SagDIG, NGC205, AndI, AndII, AndIII, AndV, AndVII, AndXI, AndXII, AndXIII, AndXIV, AndXV, AndXVI, AndXVIII, AndXXI, VV124.

\subsection{Discussion of Local Group Morphologies for Associated Objects}

We now discuss whether the associated objects identified in this paper have any signatures of a past interaction with the Milky Way.  As described in Section \ref{intro}, we expect that the passage through the larger potential of the Milky Way will affect a morphological transformation of objects in the Local Group.    Indeed, recent work using SDSS has shown that quenching of galaxies with stellar mass  $M_\star < 1 \times 10^9 M_{\odot}$ does not occur beyond 1.5 Mpc of a more massive galaxy (like M31 or the Milky Way).  This is strong evidence that an interaction with a massive galaxy is necessary for quenching \citep{Geha12}, and by extension, that galaxies which have interacted with a Milky Way-like object, can carry a morphological signature of that interaction, and be found out to 1.5Mpc, which is the same distance range found in this paper.

Possible signatures of association include low gas mass fraction due to gas stripping, a dynamically heated old population of stars, a barred or spheroidal stellar component due to tidal stirring, and a star formation history that is bursty due to gas inflows or starvation.  The strength of these transformation signatures depends on both the duration and depth of any pericentric encounter with the Milky Way, the mass of the dwarf, and to a lesser extent, whether it is a member of an infalling group.  While tidal effects scale with the relative densities of the parent and satellite galaxies (and hence are not necessarily mass-dependent), the importance of shock heating and ram-pressure stripping of gas does depend on the depth of the satellite's potential well. 
 
The similarity in the distributions of backsplash and weakly associated haloes in VLII suggests there is no easy way  to assess the nature of pericentric passages from the locations and velocities of field dwarfs. However, we do have information on their masses.  Moving from most to least massive, the 10 objects which have greater than 50 per cent likelihood of association with the Milky Way, are: irregulars, dwarf irregulars, a dwarf elliptical/spheroidal, 'transition' objects, and  dwarf spheroidals.  Since transformations are stronger in smaller galaxies, we might expect that the effects of a passage through the Milky Way could have resulted in just this sequence in morphologies.

\begin{figure}
\begin{center}
\includegraphics[width=84mm]{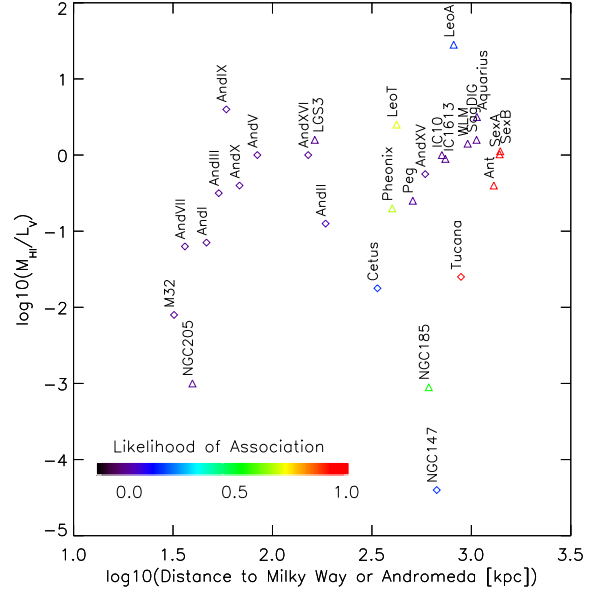}
\caption{Detected HI masses over total masses with distance to the Milky Way in kpc.  Red stars represent objects which are associated with the Milky Way. Black diamonds are field objects, and blue diamonds are satellites of Andromeda.  HI mass fraction as a function of distance is lower for associated haloes than field objects and lowest for satellites.}
\label{fig.grc}
\end{center}
\end{figure}

We use gas detections from \citet{Grcevich09} to create Figure \ref{fig.grc}, which shows detected HI mass fraction vs. distance to MW or M31 (see their Fig.3), but also includes in a colour coding the likelihood of association with the MW. From this figure it is apparent that HI gas fractions for objects with a high likelihood of association are lower than those for field objects at a given distance from the Milky Way.  This trend supports our findings, and provides further evidence that associated objects may have been stripped during their passage past the Milky Way.

Finally, there are cases where we also see hints of past interactions in the stellar populations of these objects. Most obviously Tucana and Cetus both have an old population, with no contributions from younger stars, presumably because star formation was truncated as gas was stripped during the encounter.  Antlia, NGC6822, Leo T, NGC 185 and Phoenix have all have extended old haloes, no population of intermediate-age stars, and a dynamically cold, young core \citep{Hwang11,McQuinn10}.  In these cases, the encounter could have stripped gas to delay any ongoing star formation and heated the old population.  Subsequent re-accretion of gas (or retention of a small amount of gas), funnelled to the centre by residual tidal distortions, could have formed the young population.

Overall, we conclude that these combined morphological, gas content and stellar populations signatures suggest that some, if not all of the objects we identify as ``associated'' indeed had some past interaction with the Milky Way.

\section{Summary of Major Results}\label{sec.summ}

We demonstrate that with just the line-of-sight distance and velocity, we can obtain a rough interaction history for field objects in the Local Group via comparison with VLII populations.  We separate field haloes in VLII into categories:  associated haloes have been within the virial radius of the main Milky Way-like halo, unassociated haloes have not.  

We find  $\sim$13 per cent of field haloes in the simulations to have passed through the virial volume of the Milky Way-like halo at some point during their histories.  These associated haloes could be found out to 5 \Rvir.  This suggests that, for the Local Group, of the 54 known galaxies within this distance range, we expect at least 7 to have interacted with the Milky Way.  
Further analysis of VLII suggest that these associated objects are likely to have positive radial velocities with respect to the Milky Way of order or greater than the Hubble Flow, which will make them distinguishable from the unassociated populations.  From our analysis we do not expect a mass-distance bias in the associated dwarfs around the Milky Way.  About 4 per cent of the MW-associated haloes may have become renegade haloes bound to M31.

The separation between the associated and unassociated populations in the distance-velocity plane in VLII was applied in the Local Group to identify field dwarfs that may be associated with the Milky Way:  Tucana, Cetus, Antlia, NGC3109, SextansA, SextansB, NGC6822, Phoenix, LeoT and NGC185.  Several of these objects have signatures in their morphology, gas content, or stellar populations that could be the result of their passage through the Milky Way.  This possibility should be considered when analyzing transformative internal and external effects for these objects.  Overall we conclude that our simple test provides strong support for scenarios in which the gas-poor, dwarf spheroidal objects in the field result from the transformation of gas-rich irregulars during past interactions with Milky Way or Andromeda.

\section*{Acknowledgements}
MT would like to thank Jana Grcevich for her valuable insights. MK and KVJ thanks the KITP in Santa Barbara for providing great hospitality and a stimulating environment during the First Galaxies and Faint Dwarfs conference and program, in which part of this work was completed. MKs contributions were supported in part by the National Science Foundation under Grant No. NSF PHY05-51164, OIA-1124453 (PI P.~Madau), and OIA-1124403 (PI A.~Szalay).

\end{document}